\documentclass[11pt,twoside]{article}
\usepackage{./asp2014}
\usepackage{graphicx}

\aspSuppressVolSlug
\resetcounters

\begin{document}

\title{Calibration of the Instrumental Crosstalk for the Near-IR Imaging Spectropolarimeter at the NST}
\author{Kwangsu Ahn,$^1$ and Wenda Cao$^1$}
\affil{$^1$Big Bear Solar Observatory, 40386 North Shore Lane, Big Bear City, CA, 92314, USA; \email{ksahn@bbso.njit.edu}}

% This section is for ADS Processing.  There must be one line per author.
\paperauthor{Kwangsu Ahn}{ksahn@bbso.njit.edu}{ORCID_Or_Blank}{Big Bear Solar Observatory}{Physics}{Big Bear City}{CA}{92314}{USA}
\paperauthor{Wenda Cao}{ksahn@bbso.njit.edu}{ORCID_Or_Blank}{Big Bear Solar Observatory}{Physics}{Big Bear City}{CA}{92314}{USA}

\begin{abstract}
The Near-IR Imaging Spectropolarimeter (NIRIS) is a polarimeter that is installed at the New Solar Telescope at Big Bear Solar Observatory. This instrument takes advantages of the highest spatial resolution and flux. The primary mirror is an off-axis type, so it was our interest to evaluate its contribution to the crosstalk among the Stokes parameters since we could not put our calibration optics before the mirror. We would like to present our efforts to compensate for the crosstalk among Stokes profiles caused by the relay optics from the telescope to the detector. The overall data processing pipeline is also introduced.
\end{abstract}

\section{Introduction}
The Sun's polarization signals are represented in the Stokes parameter forms, I, Q, U, and V. They all represent intensity, vertical and horizontal linear polarization, diagonal polarization, and circular polarization, respectively. As there is more sophisticated optical train used for new telescopes, all the combination of the optical elements also create polarization signals. In the case of the New Solar Telescope (NST) in Big Bear, most of the instruments sit in the Coud\'e laboratory. The sunlight from the telescope is fed down to this room, through its equatorial axis. There are about 20 mirror elements in front of the detector of each instrument. Also, the incoming beam from the telescope rotates through the hour angle and declination of the telescope pointing. Time-wise image rotation through the set of relay optics creates a considerable influence on the incoming polarization signal, where calibration of the instrumental crosstalk is an essential procedure.

In this study, we would like to share the effort to calibrate instrumental crosstalk for the NIRIS. We will also discuss extra efforts under the particular circumstances of its hardware (camera, modulator). We hope it could be helpful for those who consider similar types of the polarimeter.

\section{Instrument}

The primary and secondary mirrors of the NST have off-axis paraboloid and ellipsoid shapes, respectively. The major drawback of the off-axis nature is that these mirrors can create artificial polarization signal due to the asymmetry of the mirror structures. The challenge is that there is no linear polarizer nor quarter-wave plate that could cover the entire aperture of the primary mirror, thus leading to uncorrectable errors in polarimetry.

We are currently operating a polarimeter in the near-infrared bands, called Near-InfraRed Imaging Spectro-polarimeter (NIRIS). Detailed specifications of the NIRIS is described in \cite{ex_0}.

We now use a 2k by 2k Teledyne array for infrared imaging. Its detector chip is a CMOS array that consists of 32 channels. Its readout occurs in a serial sequence from the first element in a channel to the last one, thus introducing a time delay in exposure among pixels. It takes about 13 msec to complete such ripple readout. This readout pattern creates extra errors in determining the polarization signal.

The optical elements for calibration of the crosstalk consist of a linear polarizer and a quarter-wave plate. The quarter-wave plate consists of birefringent polymer material, so we put an extra plate for UV cutoff. These optics stay between the secondary mirror (M2) and the third mirror (M3).

\section{Data Processing}

We typically scan 60 wavelengths with 16 frames for each wavelength in 30 seconds. All the observed data were dark subtracted and flat-fielded. We took the dark and flat files with the same exposure time as in the observation runs. The dark was made by shutting down the primary mirror cover to ensure the same stray light environment as the observation. The flats were acquired by wobbling the telescope near the disk center to smear out any feature on the Sun. Then, any hot/dead pixels are masked by adjacent pixels.

\begin{figure}[ht!]
\begin{center}
\includegraphics[width=0.45\textwidth]{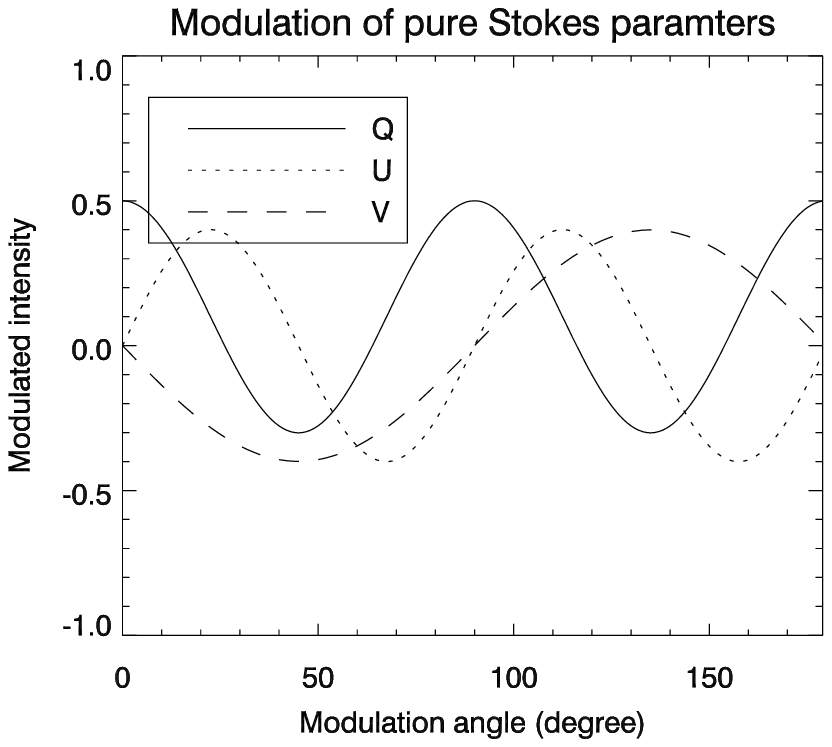}
\includegraphics[width=0.45\textwidth]{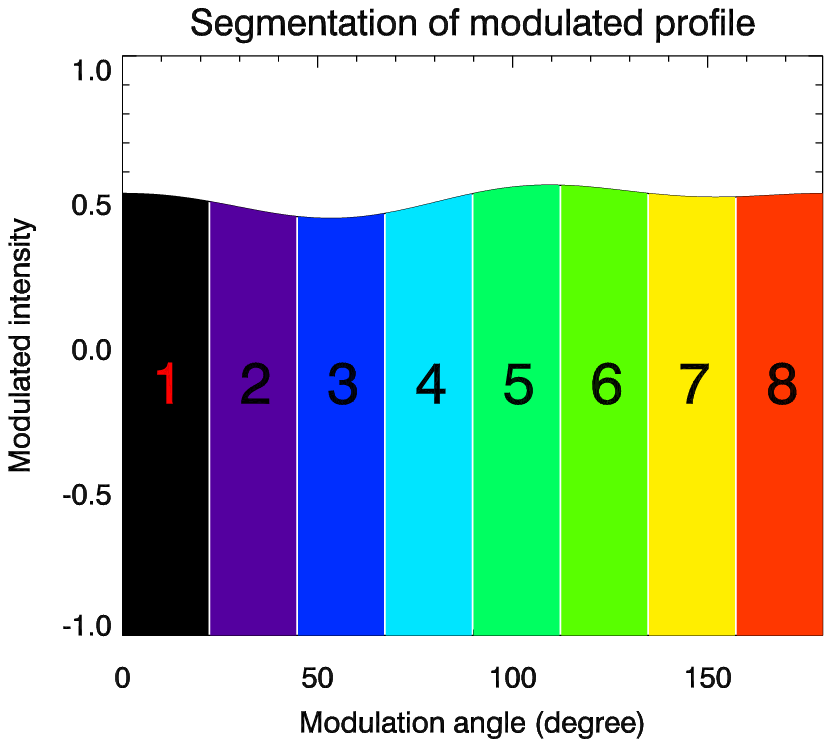}
\caption{\textit{Left:} Angular modulation pattern for each Stokes parameter, \textit{Right:} A typical modulation pattern and possible 8 integration sets for determining Stokes paramters}
\label{fig:mod}
\end{center}
\end{figure}

We use a mechanically-rotating birefringent modulator. The intensity at the detector $I''$ is a function of retardation $\delta$, modulated by sinusoidal components with respect to the fast axis angle of the modulator $\theta$. The modulation pattern is formulated as below and shown in Figure \ref{fig:mod}.

\begin{equation}
I''=\frac{1}{2}\left(I'+\frac{Q'}{2} \left( \left( I' + \cos \delta\right) +\left(1-\cos \delta \right) \cos 4\theta \right) + \frac{U'}{2} \left( 1 -\cos \delta \right) \sin 4 \theta - V' \sin \delta \sin 2\theta \right)
\end{equation}

This modulation profile is divided by 16 frames per rotation then undergoes linear combination of these to deduce 4 Stokes parameters. A set of combination for determining Stokes parameters is listed in Table 1.

\begin{table}[!ht]
\caption{List of combination of intergrations for extracting Stokes parameters}
\smallskip
\begin{center}
{\small
\begin{tabular}{cc}  % l = left, c = centered
\tableline
\noalign{\smallskip}
Parameter & Combination \\
\noalign{\smallskip}
\tableline
\noalign{\smallskip}
I & 1+2+3+4+5+6+7+8 \\
\noalign{\smallskip}
Q & 1-2-3+4+5-6-7+8 \\ % Sometimes you have empty cells
\noalign{\smallskip}
U & 1+2-3-4+5+6-7-8 \\
\noalign{\smallskip}
V & -1-2-3-4+5+6+7+8\\
\noalign{\smallskip}
\tableline\
\end{tabular}
}
\end{center}
\end{table}

The camera only offers free-run mode, where the camera's exposure timing cannot be triggered externally. Meanwhile, phase control of the modulator is challenging due to possible time delay of signal transmission from the camera to the motion controller. We can only match the rotation speed of the modulator to the exposure time of the camera. It thus results in a random phase offset between the two units. Thus, this random offset changes Stokes signals to $[I'',Q'',U'',V'']^T$ even before correcting instrumental crosstalk.

Without an offset, the combination for Stokes $Q'$ is as follows.

\begin{eqnarray}
\int_0^\frac{\pi}{8} I'' \mathrm{d}\theta
-\int_\frac{\pi}{8}^\frac{\pi}{4} I'' \mathrm{d}\theta
-\int_\frac{\pi}{4}^\frac{3\pi}{8} I'' \mathrm{d}\theta
+\int_\frac{3\pi}{8}^\frac{\pi}{2} I'' \mathrm{d}\theta \nonumber \\
+\int_\frac{\pi}{2}^\frac{5\pi}{8} I'' \mathrm{d}\theta
-\int_\frac{5\pi}{8}^\frac{3\pi}{4} I'' \mathrm{d}\theta
-\int_\frac{3\pi}{4}^\frac{7\pi}{8} I'' \mathrm{d}\theta
+\int_\frac{7\pi}{8}^\pi I'' \mathrm{d}\theta \nonumber \\ = 0.8 Q'
\end{eqnarray}

With an offset $\phi$ there is a contribution from $U'$.

\begin{eqnarray}
0.8 Q''(\phi)=\int_\phi^{\frac{\pi}{8}+\phi} I'' \mathrm{d}\theta
-\int_{\frac{\pi}{8}+\phi}^{\frac{\pi}{4}+\phi} I'' \mathrm{d}\theta
-\int_{\frac{\pi}{4}+\phi}^{\frac{3\pi}{8}+\phi} I'' \mathrm{d}\theta
+\int_{\frac{3\pi}{8}+\phi}^{\frac{\pi}{2}+\phi} I'' \mathrm{d}\theta \nonumber \\
+\int_{\frac{\pi}{2}+\phi}^{\frac{5\pi}{8}+\phi} I'' \mathrm{d}\theta
-\int_{\frac{5\pi}{8}+\phi}^{\frac{3\pi}{4}+\phi} I'' \mathrm{d}\theta
-\int_{\frac{3\pi}{4}+\phi}^{\frac{7\pi}{8}+\phi} I'' \mathrm{d}\theta
+\int_{\frac{7\pi}{8}+\phi}^{\pi+\phi} I'' \mathrm{d}\theta \nonumber \\= 0.8 (Q' \cos 4\phi - U' \sin 4 \phi )
\end{eqnarray}

This offset reduces signal in $V'$.

\begin{eqnarray}
%-\int_\phi^{\frac{\pi}{8}+\phi} I'' \mathrm{d}\theta
%-\int_{\frac{\pi}{8}+\phi}^{\frac{\pi}{4}+\phi} I'' \mathrm{d}\theta
%-\int_{\frac{\pi}{4}+\phi}^{\frac{3\pi}{8}+\phi} I'' \mathrm{d}\theta
%-\int_{\frac{3\pi}{8}+\phi}^{\frac{\pi}{2}+\phi} I'' \mathrm{d}\theta \nonumber \\
%+\int_{\frac{\pi}{2}+\phi}^{\frac{5\pi}{8}+\phi} I'' \mathrm{d}\theta
%+\int_{\frac{5\pi}{8}+\phi}^{\frac{3\pi}{4}+\phi} I'' \mathrm{d}\theta
%+\int_{\frac{3\pi}{4}+\phi}^{\frac{7\pi}{8}+\phi} I'' \mathrm{d}\theta
%+\int_{\frac{7\pi}{8}+\phi}^{\pi+\phi} I'' \mathrm{d}\theta \nonumber \\
0.8 V''(\phi)= 0.8 V' \cos 2 \phi
\end{eqnarray}

This offset $\phi$ can be corrected by multiplying a restoration matrix to the measured Stokes vector, yielding the intended Stokes vector before calibration of instrumental crosstalk, $[I', Q', U', V']^T$.

\begin{equation}
\begin{bmatrix}
I' \\
Q' \\
U' \\
V'
\end{bmatrix}
=
\begin{bmatrix}
1 & 0 & 0 & 0 \\
0 & \cos4\phi & \sin4\phi & 0 \\
0 & -\sin4\phi & \cos4\phi & 0 \\
0 & 0 & 0 & \sec2\phi
\end{bmatrix}
\begin{bmatrix}
I''\\
Q''\\
U''\\
V''
\end{bmatrix}
\end{equation}

We performed the alignment of images in several steps. First, we aligned all the frames in a scan sequence with the first frame of the first wavelength as the reference image. We only used running reference images near the line center, where the contrast of granule pattern reverse. Next, we calculated the difference of optical aberration pattern from the dual beams $A$ and $B$ by averaging the distortion patterns from 16 frame images. Since the dual beams are taken at the same time, there is only optical aberration effect between $A$ and $B$.

The calibration of instrumental crosstalk was performed by expanding a day of calibration measurement into the whole year. For this, we needed to model our optical elements by grouping them among coordinate breaks. The original Stokes parameters $[I, Q, U, V]^T$ should be deduced by multiplying the Mueller matrix of the optical train $\textbf{M}$ to the Stokes vector before calibration.

\begin{equation}
\begin{bmatrix}
I\\
Q\\
U\\
V
\end{bmatrix}
= \textbf{M}^{-1}
\begin{bmatrix}
I'\\
Q'\\
U'\\
V'
\end{bmatrix}
\end{equation}

The Mueller matirx $\textbf{M}$ is a combination of each optical element, starting from the telescope $\textbf{T}$.
\begin{equation}
\textbf{M}=\textbf{M}_n\textbf{M}_{n-1}\textbf{M}_{n-2}\dots \textbf{M}_3\textbf{T}
\end{equation}

Each mirror element can function as a linear polarizer and a retarder \citep{ex_1}. $r_s/r_p$ is a ratio of reflectivity at perpendicular linear polarization directions with respect to its fast axis, while $\delta$ is a retardation of the surface.

\begin{equation}
\textbf{M}=
\begin{bmatrix}
1 & \frac{1-\frac{r_s}{r_p}}{1+\frac{r_s}{r_p}} & 0 & 0 \\
\frac{1-\frac{r_s}{r_p}}{1+\frac{r_s}{r_p}} & 1 & 0 & 0 \\
0 & 0 & \frac{2\sqrt{\frac{r_s}{r_p}}\cos\delta}{{1+\frac{r_s}{r_p}}} & \frac{2\sqrt{\frac{r_s}{r_p}}\sin\delta}{{1+\frac{r_s}{r_p}}} \\
0 & 0 & -\frac{2\sqrt{\frac{r_s}{r_p}}\sin\delta}{{1+\frac{r_s}{r_p}}} & \frac{2\sqrt{\frac{r_s}{r_p}}\cos\delta}{{1+\frac{r_s}{r_p}}}
\end{bmatrix}
\end{equation}

The Mueller matrix is a function of declination and hour angle, which involves coordinate rotation matrix $\textbf{R}$.

\begin{equation}
\textbf{M}(dec,HA)= \textbf{M}_{rest}\textbf{R}(HA)\textbf{M}_4\textbf{R}(dec)\textbf{M}_3\textbf{R}(\phi_1)\textbf{T}
\end{equation}

Now, our problem is reduced to obtaining the constant $\textbf{M}_{rest}$ from the measured data points at different declinations and hour angles. We assumed that the contribution of the $\textbf{T}$ is within our calibration error.

\begin{equation}
 \textbf{M}_{rest}\simeq\textbf{M}\left[\textbf{R}(HA)\textbf{M}_4\textbf{R}(dec)\textbf{M}_3\textbf{R}(\phi_1)\right]^{-1}=constant
\end{equation}

\section{Results}
\begin{figure}[ht!]
\begin{center}
\includegraphics[width=0.45\textwidth]{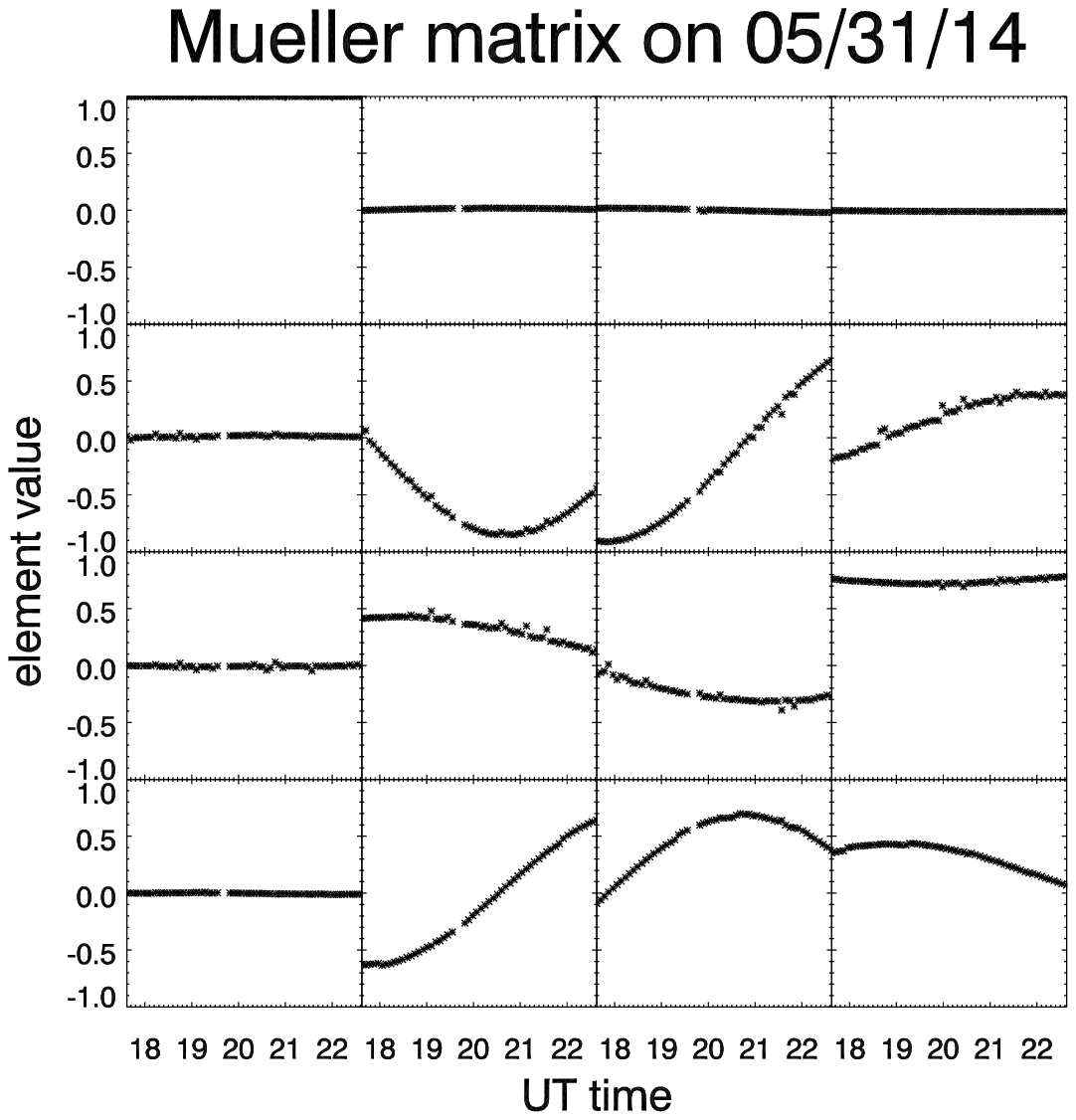}
\includegraphics[width=0.45\textwidth]{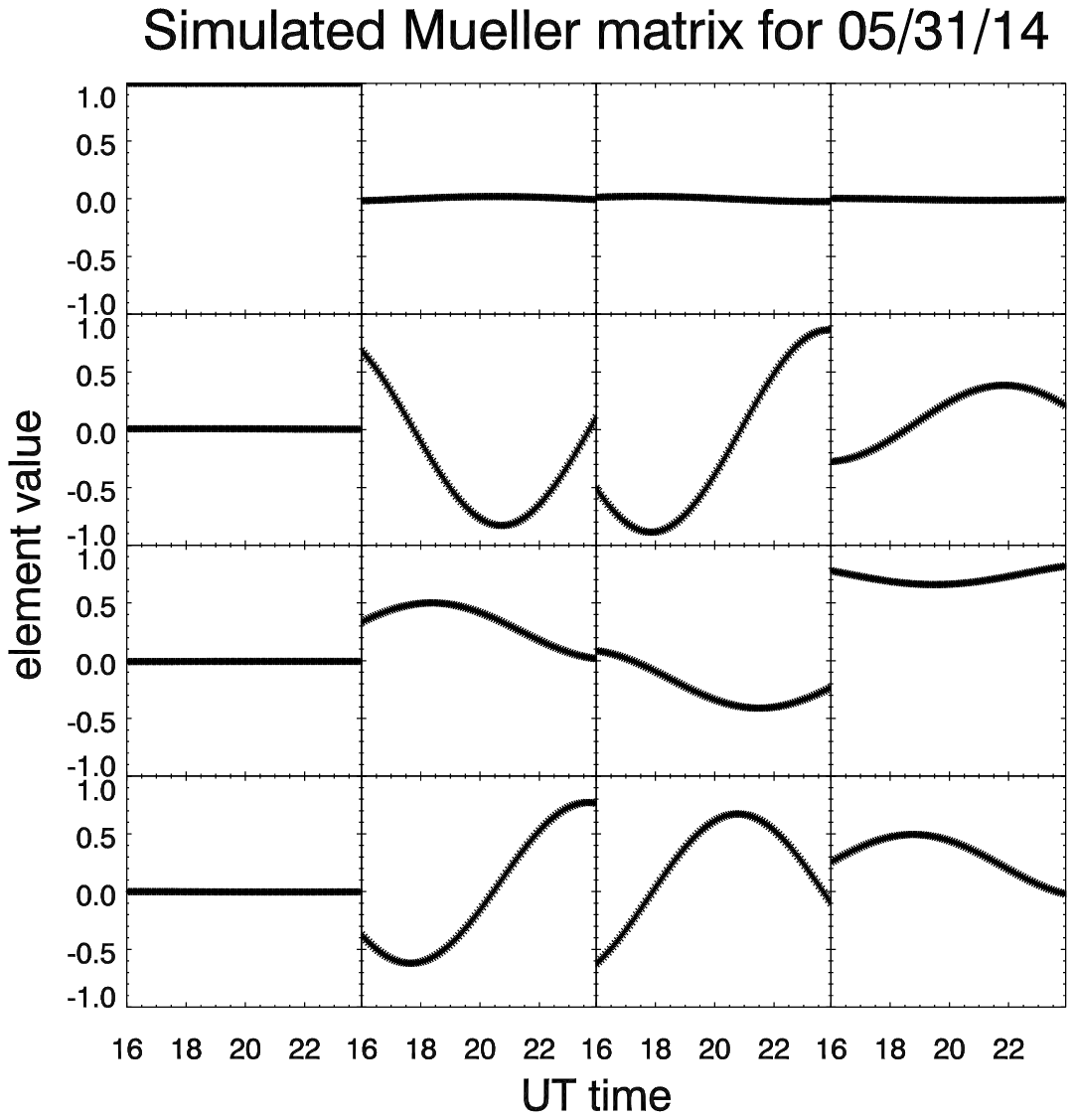}
\includegraphics[width=0.45\textwidth]{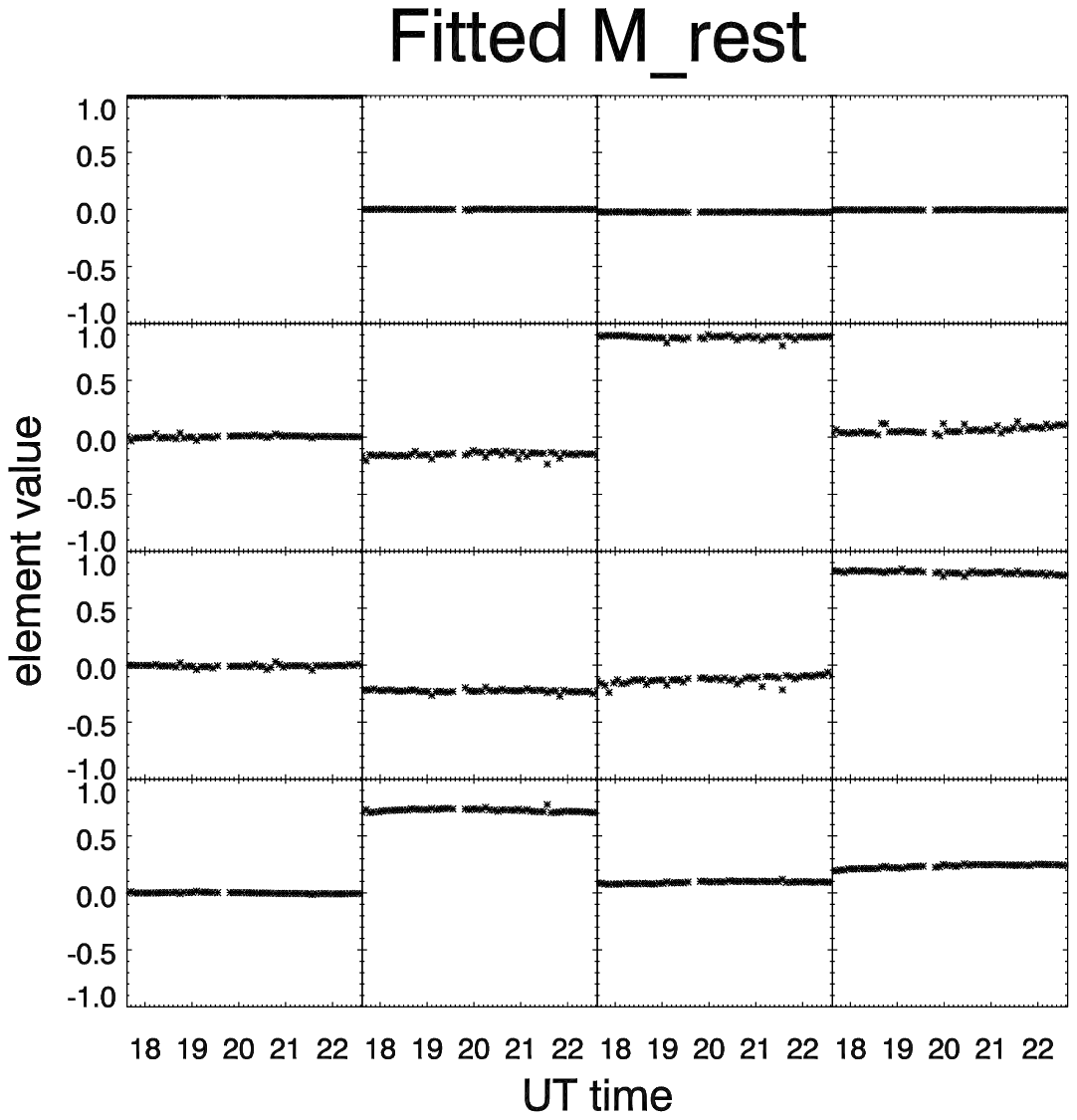}
\caption{\textit{Left:} Measured Mueller matrix pattern observed on May 31, 2014, \textit{Right:} Simulated Mueller matrix pattern for May 31, 2014, derived from the calibration measurement on October 12, 2014, \textit{Bottom:} Fitted $\textbf{M}_{rest}$ profiles from the best parameter combination of $\textbf{M}_{3}$ and $\textbf{M}_{4}$}
\label{fig:sim}
\end{center}
\end{figure}

The measured crosstalk over a course of a day on May 31, 2014 is depicted on the left panel of Figure \ref{fig:sim}. Most of the matrix elements are showing sinusoidal changes with a period of 12 hrs. The incoming light to the optical train experiences rotations as the telescope changes its geometry over time. Generally, we took a sample set of calibrated observation every 6 mins, so we could have about 60 Mueller matrix data points for one-day observation of 6 hrs. Now after applying inverse matrices of coordinate breaks from declinations and hour angles, we tried every possible combination of $r_s/r_p$ and $\delta$ for $\textbf{M}_3$ and $\textbf{M}_4$ that minimized the standard deviation of $\textbf{M}_{rest}$ from the measured profiles. The bottom panel in Figure \ref{fig:sim} shows the $\textbf{M}_{rest}$ profiles after applying the best combination of $\textbf{M}_3$ and $\textbf{M}_4$ parameters. The fitted $r_s/r_p$ and $\delta$ values for $\textbf{M}_3$ and $\textbf{M}_4$ are as below.

\begin{itemize}
\item r$_s$/r$_p$ = -0.004 for both $\textbf{M}_3$ and $\textbf{M}_4$
\item $\delta$$_{M3}$=-25$^\circ$
\item $\delta$$_{M4}$=-27$^\circ$
\end{itemize}

Since M3 and M4 are coated with the same material (protected Aluminum) and are made nearly at the same time, it is reasonable to have similar or same parameter values.

Once we have $\textbf{M}_{rest}$ determined from the measurement, now we can simulate the Mueller matrix of the whole system $\textbf{M}$ by multiplying coordinate breaks for each declination and hour angles. Right panel in Figure \ref{fig:sim} is a simulated Mueller matrix for May 31, 2014 derived from a separate calibration measurement on Oct 12, 2014, and they match well.

After calibration of the instrumental crosstalk, we estimated the remaining crosstalk among $Q$, $U$, and $V$. Our analysis showed that such residual crosstalk remained up to 10 percent of the original signals, where the influence of the primary and secondary mirrors could play a role. Thus, we performed the second order calibration, which subtracts a portion of Stokes $V$ map from $Q$ and $U$ to minimize the degree of anti-symmetric line profiles in $Q$ and $U$. We also calculated the partial combination of $Q$ and $U$ map that minimize the symmetric profiles in $V$. This final step was essential to perform reliable Milne-Eddington inversion process.

\section{Discussion and Conclusion}
We have shown that it is possible to perform Stokes map calibration with the one-day observation of calibration measurement. This was possible by an assumption that the properties of the optics do not change over time. In this manner, do not need to sacrifice much observing time to use for calibration purpose.

It may not be always true, however, to assume that the optical properties do not change. Even though we assumed that the oxidized coating is already mature, there may be still oxidization layer slowly building up. Any imperfect alignment of the relay optics, or wobble of the incoming beam will contribute to the time-dependent variation of the $\textbf{M}_{rest}$ properties. When there is any change of mirror then previous simulation may not be valid any longer.
 Therefore, even though a single day calibration measurement with simulation should perform well for a while, there still is a need to perform repeated calibration measurements. In our case, calibration measurement of 2 times a year or more would be a safe interval. In this manner, we can compare a measured Mueller matrix for a day to the simulated ones for the day, derived from another calibration measurements.

% For non-BibTex:

\end{document}